\documentclass[aps,twocolumn,pra,showpacs]{revtex4}
\usepackage{exscale}
\usepackage{graphicx}
\usepackage{amsmath}
\usepackage{latexsym}
\usepackage[normalem]{ulem}
\usepackage{amsfonts}
\usepackage{amssymb}
\usepackage{amscd}
\usepackage{bbm}
\usepackage{dcolumn}      
\usepackage{bm}           
\usepackage{pstricks,pst-grad,fancybox,graphics}
\usepackage{enumerate}

\newcommand{\bra}[1]{\ensuremath{\langle#1|}}

\newcommand{\ket}[1]{\ensuremath{|#1\rangle}}

\newcommand{\ketbra}[1]{\ensuremath{| #1 \rangle \langle #1 |}}

\newcommand{\BE}{\begin{equation}}
\newcommand{\EE}{\end{equation}}
\newcommand{\be}{\begin{equation}}
\newcommand{\ee}{\end{equation}}
\newcommand{\bea}{\begin{eqnarray}}
\newcommand{\eea}{\end{eqnarray}}
\newcommand{\bean}{\begin{eqnarray*}}
\newcommand{\eean}{\end{eqnarray*}}
\newcommand{\kommentar}[1]{}

\newcommand{\mean}[1]{\ensuremath{\langle #1 \rangle}}

\newcommand{\proj}[1]{\ketbra{#1}}
\newcommand{\tr}{{\rm Tr}}
\newcommand{\diag}{{\rm diag}}

\newcommand{\bc}{\begin{center}}
\newcommand{\ec}{\end{center}}
\newcommand{\proofend}{\hfill\fbox\\\medskip }

\begin{document}

\title{Fisher information and multiparticle entanglement}

\author{Philipp Hyllus$^{1,2}$, 
Wies{\l}aw Laskowski$^{3,4,5}$,
Roland Krischek$^{4,5}$, 
Christian Schwemmer$^{4,5}$, 
Witlef Wieczorek$^{4,5,6}$, 
Harald Weinfurter$^{4,5}$,
Luca Pezz{\'e}$^7$, 
and
Augusto Smerzi$^{1,7}$, 
}
\affiliation{
$^1$INO-CNR BEC Center and Dipartimento di Fisica, Universit{\`a} di Trento, I-38123 Povo, Italy\\
$^2$Department of Theoretical Physics, The University of the Basque Country, P.O. Box 644, E-48080 Bilbao, Spain\\
$^3$Institute of Theoretical Physics and Astrophysics, University of Gda\'nsk, PL-80-952 Gda\'nsk, Poland\\
$^4$Fakult\"at f\"ur Physik, Ludwig-Maximilians Universit\"at M\"unchen, D-80799 M\"unchen, Germany\\
$^5$Max-Planck Institut f\"ur Quantenoptik, D-85748 Garching, Germany\\
$^6$Present address: Vienna Center for Quantum Science and Technology (VCQ), Faculty of Physics,
University of Vienna, Boltzmanngasse 5, 1090 Vienna, Austria\\
$^7$INO-CNR and LENS, Largo Fermi 6, I-50125 Firenze, Italy}



\date{\today}

\begin{abstract}
The Fisher information $F$ gives a limit to the ultimate precision 
achievable in a phase estimation protocol. 
It has been shown recently that the Fisher information for 
linear two-mode interferometer cannot exceed the number of particles 
if the input state is separable. 
As a direct consequence, with such input states the shot-noise limit is the 
ultimate limit of precision. In this work, we go a step further by 
deducing bounds on $F$ for several multiparticle entanglement classes. 
These bounds imply that genuine multiparticle entanglement is needed
for reaching the highest sensitivities in quantum interferometry.
We further compute similar bounds on the average Fisher information $\overline F$
for collective spin operators, where the average is performed over all possible
spin directions. 
We show that these criteria detect different sets of states and illustrate their
strengths by considering several examples, also using experimental
data. In particular, the criterion based 
on $\overline F$ is able to detect certain bound entangled states.
\end{abstract}

\maketitle

\section{Introduction}
\label{sec:intro}

Entanglement is a distinguishing feature of quantum theory
and will play a key role in the development of future technologies.
Indeed, by using many-particle entangled states it is possible 
to perform several tasks better than feasible with 
any classical means \cite{NielsenBook}.
A valuable example is the estimation of a phase shift $\theta$ as done in quantum 
interferometry \cite{GiovannettiSci04,GiovannettiNatPhot11,WisemanBook10}.
In this case, by using a probe state of $N$ classically correlated particles  
it is possible to reach, at maximum, a phase uncertainty which 
scales as $\Delta \theta \sim 1/\sqrt{N}$.
This bound, generally indicated as the shot noise limit, 
is not fundamental and can be surpassed by preparing the $N$
particles in a proper entangled state.
It is therefore important to have a precise classification of 
entangled states and study their usefulness for specific applications.

While the structure of the set of entangled bipartite quantum states is 
understood quite well, less is known about the classification 
and quantification of the entanglement of multipartite quantum states
\cite{PlenioQIC07,AmicoRMP08,HorodeckiRMP09,GuehnePR09}. 
Commonly applied criteria to distinguish between different
entanglement classes include entanglement witnesses 
\cite{BourennanePRL04,KaszlikowskiNJP08,KrammerPRL09,BancalPRL11},
criteria inspired by or derived from Bell inequalities 
\cite{SvetlichnyPRD87,GisinPLA98,CollinsPRL02,SeevinckPRA01,NagataPRL02,YuPRL03,
LaskowskiPRA05,SchmidPRL08,BancalPRL09},
and spin-squeezing inequalities 
\cite{SoerensenPRL01,DurkinPRL05,VitaglianoPRL11,DuanPRL11}.
Recently, other approaches have led to criteria which can be
evaluated directly from elements of the density matrix
\cite{GuehneNJP10,HuberPRL10}. 
Further recent work on the detection of 
multiparticle entanglement can be found in the Refs
\cite{LiPRL10,JungnitschPRL11,deVicentePRA11,HuberPRA(R)11}
and in the recent review Ref.~\cite{GuehnePR09}.

In this manuscript, we introduce novel criteria
which can distinguish between different multipartite entanglement classes
and which are deeply connected to phase estimation. This extends
previous works 
\cite{SoerensenNat01,SoerensenPRL01,GiovannettiPRL06,PezzePRL09,HyllusPRA10,HyllusPRL10}
on the interplay between entanglement and phase sensitivity.
Our criteria are based on the quantum Fisher information (QFI) for linear two-mode transformations
and can be easily computed for any density matrix $\rho$ of an arbitrary number of particles.
The first set of criteria is obtained by optimizing
the QFI for different multipartite entanglement classes.
We discuss bounds on the QFI that can be beaten 
only by increasing the number of entangled particles in the probe state.
Our classification distinguishes quantum phase
estimation in the sense that genuine multiparticle entanglement
is necessary to accomplish this quantum task in the best possible way. 
The second set of criteria is based on the QFI 
for linear collective spin operators, averaged over all
spin directions in the Bloch sphere.
The sets of states detected by the two
criteria are different and not contained in each other.
We consider several examples in order to assess the strength
of the criteria. In particular, using experimental data we apply
our criteria for several states of $N=4$ photons.

The article is organized as follows. We start by 
introducing the basic concepts related to general phase estimation 
protocols, linear two-mode interferometers, and the classification of
multiparticle entanglement in Section~\ref{sec:basics}.
In Section~\ref{sec:criteria} we derive and compare the entanglement 
criteria based on the QFI
and on the average QFI. 
In Section~\ref{sec:examples}, we apply the criteria to several 
families of entangled states, including experimental data.
We conclude in Section~\ref{sec:conclusions}.

\section{Basic concepts}
\label{sec:basics}

\subsection{Phase Estimation and Entanglement}

In a general phase estimation scenario, a probe state $\rho$ 
is transformed into $\rho(\theta) = e^{-i \theta \hat H} \, \rho \, e^{+i \theta \hat H}$, 
depending on the (unknown) phase shift $\theta$ and the operator $\hat H$.
The phase shift is inferred as the value assumed by an estimator,
$\theta_{\rm est}(\{\mu_i\}_m)$, depending on the results 
$\{\mu_i\}_m=\{\mu_1, ...,\mu_m\}$ of $m$ independent repeated measurements of 
a positive operator valued measurement (POVM) with elements $\{\hat E_\mu\}_\mu$.
We indicate with $\mean{\theta_{\rm est}}$ and 
$(\Delta \theta_{\rm est})^2 = \mean{\theta_{\rm est}^2} - \mean{\theta_{\rm est}}^2$
the mean value and variance of the estimator, respectively, 
calculated over all possible sequences $\{\mu_i\}_m$.
If the estimator is unbiased, {\em i.e.} its mean value 
coincides with the true value of the phase shift, $\mean{\theta_{\rm est}}=\theta$,
then its minimal standard deviation is limited by 
the bounds \cite{Helstrom76,Holevo82}
\be \label{CR}
	\Delta \theta_{\rm{est}} \ge \frac{1}{\sqrt{ m F}} \geq \frac{1}{\sqrt{ m F_Q}},
\ee
The quantity $F$ in the first inequality is the Fisher information, defined as
\be
	F=\sum_{\mu} \frac{1}{P(\mu|\theta)} [\partial_\theta P(\mu|\theta)]^2,
\ee
where $P(\mu|\theta)=\tr[\rho(\theta)\hat E_\mu]$ are conditional probabilities.
The maximum likelihood estimator is an example of an estimator which is unbiased and 
saturates $\Delta \theta_{\rm{est}} = 1/\sqrt{ m F}$ in the central limit, for a sufficiently large $m$ \cite{maxlik}. 
According to Eq.~(\ref{CR}), $F$ thus quantifies the asymptotic usefulness of a quantum state
for phase estimation, given the operator $\hat H$ and the chosen final measurement.
 Maximizing $F$ over all possible 
POVMs leads to the so-called quantum Fisher information $F_Q$, 
and thus to the second inequality in Eq.~(\ref{CR}).
For a mixed input state $\rho=\sum_l\lambda_l\proj{l}$ (with $\lambda_l >0$, $\sum_l\lambda_l=1$)
the QFI is given by \cite{BraunsteinPRL94}
\be \label{QFI}
	F_Q[\rho;\hat H]=2\sum_{l,l'} \frac{(\lambda_{l}-\lambda_{l'})^2}{\lambda_{l}+\lambda_{l'}}
	|\bra{l}\hat H\ket{l'}|^2,
\ee
where the sum runs over indices such that $\lambda_{l}+\lambda_{l'}>0$.
For pure input states this reduces to $F_Q=4(\Delta\hat H)^2$, 
where $(\Delta\hat H)^2 = \mean{\hat H^2} - \mean{\hat H}^2$ is the variance of the
generator of the phase shift, $\hat H$ \cite{nota_FQpure}. 

In this manuscript we focus on linear two-mode interferometers and input states of $N$ particles.
In this case 
\be\label{eq:linH}
	\hat H_{\rm lin} = \frac{1}{2}\sum_{l=1}^N \hat\sigma_{\vec n_l}^{(l)},
\ee
where $\hat\sigma_{\vec n_l}^{(l)} = {\vec n_l}\cdot\vec{\hat \sigma}^{(l)} =
\alpha_l \hat \sigma_x^{(l)} + \beta_l \hat \sigma_y^{(l)} + \gamma_l \hat \sigma_z^{(l)}$
is an operator decomposed as the sum of Pauli matrices acting on the particle $l$, and
$\vec n_l \equiv (\alpha_l, \beta_l, \gamma_l)$ is a vector on the Bloch sphere
$(\alpha_l^2+\beta_l^2+\gamma_l^2=1)$.
If all local directions are the same, ${\vec n_l}={\vec n}$,
then $\hat H_{\rm lin} \equiv \hat J_{\vec n}={\vec n}\cdot \vec{\hat J}$, 
where $\vec{\hat J} \equiv \frac{1}{2}\sum_{l=1}^N \vec{\hat\sigma}^{(l)}$ 
is a collective spin operator.
The operators $\hat J_x$, $\hat J_y$, and $\hat J_z$ fulfill the commutation 
relations of angular momentum operators. 
As an example for a linear, collective, two-mode interferometer we mention
the Mach-Zehnder interferometer, whose generator is 
$\hat H_{\rm lin}=\hat J_y$ \cite{YurkePRA86}. 

For linear phase shift generators $\hat H_{\rm lin}$ as in Eq.~(\ref{eq:linH}),  
the QFI provides a direct connection between entanglement and phase uncertainty. 
We remind that a state of $N$ particles is entangled  
if it cannot be written as a separable state
$\rho_{\rm sep}=\sum_\alpha p_\alpha \bigotimes_{l=1}^N \proj{\psi_\alpha^{(l)}}$,
where $\{p_\alpha\}$ forms a probability distribution \cite{WernerPRA89}.
It has been recently shown that the QFI
for separable states and linear generators is \cite{GiovannettiPRL06,PezzePRL09}
\be \label{F_Q_bound_sep}
	F_Q[\rho_{\rm sep};\hat H_{\rm lin}]\le N.
\ee 
Taking into account Eqs~(\ref{CR}) and (\ref{F_Q_bound_sep}) 
and the definition of QFI, $F_Q \geq F$, 
we conclude that the phase uncertainty attainable with separable states is 
$\Delta\theta_{\rm{est}} \ge \Delta \theta_{\rm{SN}}$, where
\be\label{SNL}
	\Delta \theta_{\rm{SN}} = \frac{1}{\sqrt{mN}}.
\ee 
This bound holds for any linear interferometer and any final measurement
and is generally called the shot-noise limit.
It is not fundamental and can be surpassed by using proper
entangled states.
For general probe states of $N$ particles, we have \cite{GiovannettiPRL06,PezzePRL09}
\be \label{F_Q_bound_all}
	F_Q[\rho;\hat H_{\rm lin}]\le N^2,
\ee 
where the equality can only be saturated by certain maximally entangles states.
From the maximum value of the QFI we obtain the 
optimal bound for the phase uncertainty, called the Heisenberg limit, 
\be \label{HL}
	\Delta\theta_{\rm{HL}} = \frac{1}{\sqrt{m}\,N}.
\ee
We thus expect that, in order to increase the QFI and 
the sensitivity of a linear interferometer, it is necessary
to increase the number of entangled particles in 
the probe state.
The purpose of this manuscript is to quantitatively investigate this 
effect and to derive bounds on the 
QFI
for multiparticle entanglement classes. 

\subsection{Multiparticle Entanglement}

We consider the following classification of multiparticle 
entanglement from Ref.~\cite{SeevinckPRA01, GuehneNJP05, ChenPRA05} 
(see also \cite{SoerensenPRL01}; alternative classifications can be found in Refs.~\cite{DuerPRL99,AcinPRL01}).
A pure state of $N$ particles is $k$-producible if it can be written
as $\ket{\psi_{k-{\rm prod}}} = \otimes_{l=1}^M \ket{\psi_l}$,
where $\ket{\psi_l}$ is a state of $N_l \leq k$
particles (such that $\sum_{l=1}^M N_l=N$).
A state is $k$-particle entangled if 
it is $k$-producible but not $(k-1)$-producible.
Therefore, a $k$-particle entangled state can be written 
as a product $\ket{\psi_{k-{\rm ent}}} = \otimes_{l=1}^M \ket{\psi_l}$
which contains at least one state $\ket{\psi_l}$ of $N_l=k$ particles which does not factorize.
A mixed state is $k$-producible if it can be written as a mixture of 
$(k_l\le k)$-producible pure states, {\em i.e.},
$\rho_{k-{\rm prod}}=\sum_l p_l \proj{\psi_{k_l-{\rm prod}}}$, where $k_l\le k$ for all $l$.
Again, it is $k$-particle entangled if it is $k$-producible but not $(k-1)$-producible.
We denote the set of $k$-producible states by ${\cal S}_k$. 
We will later use that ${\cal S}_k$ is convex for any $k$.
Note that formally, a fully separable state is $1$-producible,
and that a decomposition of a $k<N$-particle entangled state of $N$ particles
may contain states where different sets of particles are entangled.

Let us illustrate the classification by considering states of $N=3$ particles. 
A state $\ket{\psi_{1-{\rm prod}}}=\ket{\phi}_1\otimes\ket{\varphi}_2\otimes\ket{\chi}_3$
is fully separable. 
A state $\ket{\psi_{2-{\rm ent}}}=\ket{\phi}_{12}\otimes\ket{\chi}_3$
which cannot be written as $\ket{\psi_{1-{\rm prod}}}$ 
(i.e. $\ket{\phi}_{12}$ does not factorize, $\ket{\phi}_{12} \neq \ket{\phi}_1\otimes\ket{\varphi}_2$) 
is $2$-particle entangled.
A state $\ket{\psi_{3-{\rm ent}}}$ which does not factorize is $3$-particle entangled. 

\section{Criteria for multiparticle entanglement from the quantum Fisher information}
\label{sec:criteria}

Now we are in a position to derive the desired bounds. 
We start by computing the maximum of the quantum Fisher information 
$F_Q[\rho_{k-{\rm prod}};\hat H_{\rm lin}]$ for $k$-producible
states and linear Hamiltonians $\hat H_{\rm lin}$,
including the case of collective spin operators $\hat H_{\rm lin} = \hat J_{\vec{n}}$.
Then, we derive similar bounds for the
quantum Fisher information for a generator $\hat J_{\vec n}$,
now averaged over all directions ${\vec n}$. At the end of this
section, we investigate the question whether or not the criteria
are different by comparing the sets of states they detect.

\subsection{Entanglement criterion derived from ${\bf F_Q}$}
\label{ssec:FQ}

\noindent
{\bf Observation 1 (${\bf F_Q^{k+1}}$ criterion).} 
{\em For $k$-producible states
and an arbitrary linear two-mode interferometer $\hat H_{\rm lin}$ defined
in Eq.~(\ref{eq:linH}), the quantum Fisher information is bounded by
\be
	\label{FQ_class}
	F_Q[\rho_{k-{\rm prod}};\hat H_{\rm lin}]\le s k^2 + r^2,
\ee
where $s=\lfloor \frac{N}{k}\rfloor$ is the largest integer smaller than
or equal to $\frac{N}{k}$ and $r=N-sk$.
Hence a violation of the bound~(\ref{FQ_class}) proves
$(k+1)$-particle entanglement.
The bounds are uniquely saturated by a product of $s$ GHZ states
of $k$ particles and another GHZ states of $r$ particles, 
where  \cite{nota_basis}
\be\label{eq:GHZ}
\ket{{\rm GHZ}_\nu}=\frac{1}{\sqrt{2}}(\ket{0}^{\otimes \nu}+\ket{1}^{\otimes \nu}),
\ee
known as the Greenberger-Horne-Zeilinger (GHZ) \cite{GreenbergerAJP90} or NOON \cite{LeeJMO02} state
of $\nu$ particles.}

{\em Proof.}
The basic ingredients of the derivations 
are the following: (i) The sets ${\cal S}_k$ of $k$-producible states 
are convex.
(ii) The Fisher information is convex in the states, {\em i.e.}, for any fixed 
phase transformation and any fixed output measurement the relation
$F[p\rho_1+(1-p)\rho_2]\le pF[\rho_1]+(1-p)F[\rho_2]$ holds
for $p\in[0,1]$ \cite{CohenIEEE68}. Since the quantum Fisher information is equal
to the Fisher information for a particular measurement, this holds also for $F_Q$.
(iii) It is easy to see that for a product state $\ket{\phi_A}\otimes\ket{\chi}_B$,
$(\Delta\hat H_{\rm lin}^{(AB)})^2_{\ket{\phi}_A\otimes\ket{\chi}_B}
=(\Delta\hat H_{\rm lin}^{(A)})^2_{\ket{\phi}_A}+(\Delta\hat H_{\rm lin}^{(B)})^2_{\ket{\chi}_B}$.
Here $\hat H^{(AB)}_{\rm lin}$ acts on all the particles while $\hat H_{\rm lin}^{(A)}$ acts
on the particles of $\ket{\psi}_A$ only and in analogy for $\hat H_{\rm lin}^{(B)}$.
(iv) For a state with $N$ particles, $4(\Delta\hat H_{\rm lin})^2\le N^2$ 
holds \cite{GiovannettiPRL06}. The inequality is saturated 
uniquely by the GHZ state.

It follows from (i) and (ii) that the maximum of $F_Q$ for a fixed 
Hamiltonian $\hat H_{\rm lin}$
and $k$-producible mixed states is reached on pure $k$-producible states 
$\ket{\psi_{k-{\rm prod}}}$ \cite{BoydBook}.
Therefore our task is to maximize 
$F_Q[\ket{\psi_{k-{\rm prod}}};\hat H_{\rm lin}]=4(\Delta\hat H_{\rm lin})^2_{\ket{\psi_{k-{\rm prod}}}}$
with respect to the probe state $\ket{\psi_{k-{\rm prod}}}$
and linear operator $\hat H_{\rm lin}$.
Since the local directions of $\hat H_{\rm lin}$ [Eq.~(\ref{eq:linH})]
can be changed by local unitary operations \cite{HyllusPRA10}, which do not change the
entanglement properties of the state, we can, without loss of generality, fix $\hat H_{\rm lin}=\hat J_z$.
Due to (iii) and (iv), we obtain
max$_{\ket{\psi_{k-{\rm prod}}}}(\Delta\hat J_z)^2_{\ket{\psi_{k-{\rm prod}}}}$
=max$_{\ket{\psi_{k-{\rm prod}}}}\sum_{l=1}^M(\Delta\hat J_z^{(l)})^2_{\ket{\psi_l}}$
=max$_{\{N_l\}}\frac{1}{4}\sum_{l=1}^M N_l^2$.
Since $(N_1+1)^2+(N_2-1)^2\ge N_1^2 + N_2^2$ if $N_1\ge N_2$,
the quantum Fisher information is increased by making the $N_l$ as
large as possible. 
Hence the maximum is reached by the product of $s=\lfloor \frac{N}{k}\rfloor$ 
GHZ state of $N_l=k$ particles and 
one GHZ state of $r=N-s k$ particles.
Therefore, for $k$-producible states, 
the quantum Fisher information is bounded 
by Eq.~(\ref{FQ_class}). 
\proofend

\begin{figure}[!h]
\begin{center}
\includegraphics[scale=0.58]{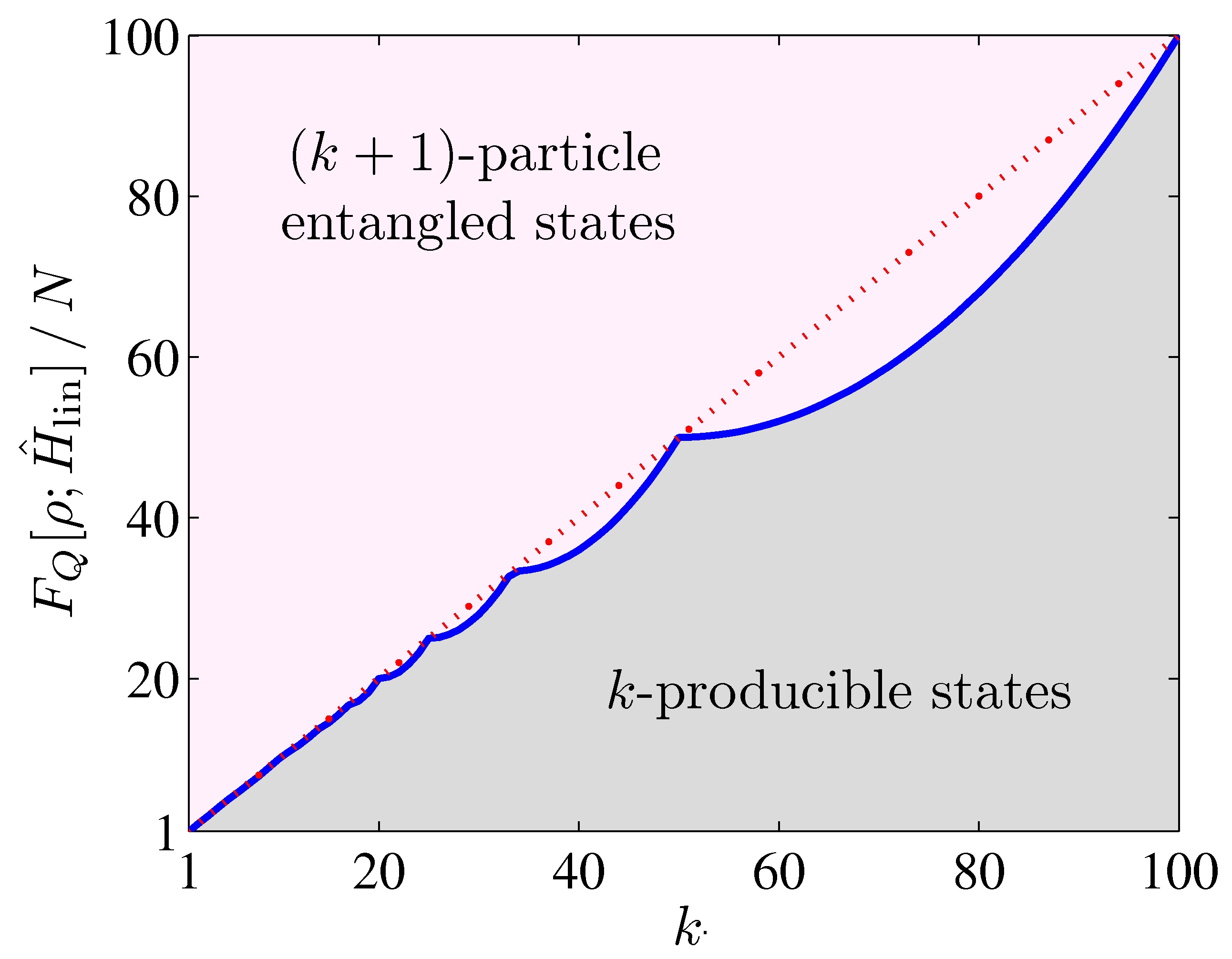}
\end{center}
\caption{\small{ $\mathbf{F_Q^{k+1}}$ {\bf criterion:} 
The solid line is the bound $F_Q[\rho;\hat H_{\rm lin}] = s k^2 + r^2$
which separates $k$-producible states (below the line)
from $(k+1)$-particle entangled states (above the line).
For comparison, the function $F_Q[\rho;\hat H_{\rm lin}]=Nk$ is plotted (dotted line). 
Here $N=100$.}}
\label{Fig1} 
\end{figure}

Given the operator $\hat H_{\rm lin}$ and the probe state $\rho$, 
the criterion~(\ref{FQ_class}) has a clear operational meaning.
If the bound is surpassed, then the probe state contains useful $(k+1)$-particle entanglement:
when used as input state of the interferometer defined by the transformation 
$e^{-i\theta\hat H_{\rm lin}}$, $\rho$ enables a phase sensitivity better than any $k$-producible state.
A plot of the bound Eq.~(\ref{FQ_class}) is presented in Fig.~\ref{Fig1} as
a function of $k$ and for $N=100$. 
Since the bound increases monotonically with $k$, 
the maximum achievable phase sensitivity increases with
the number of entangled particles.
For $k=1$ we recover the bound~(\ref{F_Q_bound_sep})
for separable states. For $k=N-1$, the bound is
$F_Q[\rho_{(N-1)-{\rm prod}};\hat H_{\rm lin}]\le (N-1)^2+1$
and a quantum Fisher information larger than this value
signals that the state is fully $N$-particle entangled.
The maximum value of the bound is obtained for $k=N$
(thus $s=1$ and $r=0$), when $F_Q[\rho_{N-{\rm ent}};\hat H_{\rm lin}] = N^2$,
saturating the equality sign in Eq.~(\ref{F_Q_bound_all}).

Given the probe state $\rho$, the $F_Q^{k+1}$ criterion
can be used to detect $(k+1)$-particle entanglement.
In order to  maximize $F_Q[\rho;\hat H_{\rm lin}]$,
it is advantageous to optimize
the local directions ${\vec n}_l$ in $\hat H_{\rm lin}$ \cite{HyllusPRA10}, see Eq.~(\ref{eq:linH}).
While the general problem needs to be solved numerically, a simple analytic solution 
can be obtained if we restrict ourselves to collective spin operators
$\hat H_{\rm lin} =\hat{J}_{\vec n}$.
In this case we have \cite{HyllusPRA10}
\be \label{eq:FQ_from_GammaC}
	F_Q[\rho;\hat J_{\vec n}]={\vec n}^T \Gamma_C {\vec n}.
\ee
The matrix $\Gamma_C$ is real and symmetric and has the entries
\be\label{eq:GammaCentries}
	[\Gamma_C]_{ij}
	=2\sum_{l,l'}\frac{(\lambda_{l}-\lambda_{l'})^2}{\lambda_{l}+\lambda_{l'}}
	{\cal R}\big[\bra{l}\hat J_{i}\ket{l'}\bra{l'}\hat J_{j}\ket{l}\big],
\ee
where the states $\ket{l}$ and the variables $\lambda_l$ are defined 
by the eigenvalue decomposition of the input state, 
$\rho=\sum_l \lambda_l \proj{l}$, and ${\cal R}(z)$ is 
the real part of $z$. The sum runs over indices
where $\lambda_{l}+\lambda_{l'}>0$.
Maximizing Eq.~(\ref{eq:FQ_from_GammaC}) with respect to $\vec n$ leads to
\be \label{FQmax}
F_Q^{\rm max}[\rho]\equiv \max_{\vec n}F_Q[\rho;\hat J_{\vec n}]=\lambda_{\rm max}(\Gamma_C),
\ee
where $\lambda_{\rm max}(\Gamma_C)$ is the maximal eigenvalue
of $\Gamma_C$. 
In the case of collective operations, the criteria Eq.~(\ref{FQ_class})
can be substituted by
\be
\lambda_{\rm max}(\Gamma_C) \leq s k^2 + r^2.
\ee
For any pure symmetric input state this is the optimal value 
of $F$ also if arbitrary local unitary operations can be used \cite{HyllusPRA10}.
Note that while this optimization increases $F_Q$, it might happen that
for a fixed output measurement the Fisher information $F$ is actually reduced
by this optimization because the measurement would have to be adapted as well
\cite{BraunsteinPRL94}.

Finally note that the result Eq.~(\ref{FQ_class}) can be obtained
directly by using the Wigner-Yanase information $I$ \cite{WignerPNAS63}.
The bound~(\ref{FQ_class}) has been derived previously for 
$I$ in Ref.~\cite{ChenPRA05}, and directly applies to the quantum 
Fisher information since $I$ is convex in the states and agrees 
with the Fisher information on pure states, 
$F[\ket{\psi};\hat H]=4I(\ket{\psi},\hat H)$.
See Ref.~\cite{TothArXiv11} for a more general discussion of convex
quantities which are equal to the Fisher information on the pure states.
Note also that a bound similar to
Eq.~(\ref{FQ_class}) has been discussed for the class of so-called
spin-squeezed states \cite{SoerensenPRL01}.

\subsection{Entanglement Criterion derived from ${\bf \overline F_Q}$ }
\label{ssec:Fav}

Let us now consider the estimation of a fixed (unknown) 
phase shift $\theta$ with an interferometer that,
in each run of the experiment, is given by $\exp[-i\hat J_{\vec \nu}\theta]$ 
with a random direction ${\vec \nu}$ of probability $P(\vec{\nu})$. 
For $m\gg 1$ independent repetitions of the phase
measurement, 
the 
phase estimation uncertainty approaches
\be
\Delta \theta_{\rm est} \geq \frac{1}{\sqrt{m F^{P}[\rho]}},
\ee
where
\be\label{eq:Fav1}
	F^{P}[\rho]=\int_{|{\vec \nu}|^2=1}{\rm d}^3{\vec \nu}\ P(\vec{\nu}) \, F[\rho;\hat J_{\vec \nu};\{\hat E_\mu\}],
\ee
and $P(\vec{\nu})$ is normalized to one.
The direction-averaged Fisher information, Eq.~(\ref{eq:Fav1}), is bounded
by 
\be\label{eq:FQav1}
	F^{P}_Q[\rho]=\int_{|{\vec \nu}|^2=1}{\rm d}^3{\vec \nu}\ P(\vec{\nu}) \, F_Q[\rho;\hat J_{\vec \nu}].
\ee
The latter quantity can be used to 
introduce an infinite set of multiparticle entanglement criteria, depending on the function
$P(\vec{\nu})$. 
If $P(\vec{\nu}) = \delta_{\vec \nu, \vec n}$, then we recover the standard situation of a fixed collective spin
direction and the criteria Eq.~(\ref{FQmax}).
We here consider the opposite case $P(\vec{\nu}) = 1/4\pi$
where all directions $\vec \nu$ on the Bloch sphere appear with equal probability.
We indicate the corresponding average of the quantum Fisher information as $\overline F_Q[\rho]$.
It can be written as 
$\overline F_Q[\rho]=\frac{1}{4\pi}\sum_{ij} [\Gamma_C]_{ij} \int_{|{\vec \nu}|^2=1}{\rm d}^3{\vec \nu}\ \nu_i \nu_j$ [see Eq.~(\ref{eq:FQ_from_GammaC})]. 
Evaluating the integrals leads to 
\be\label{eq:Fav2}
	\overline F_Q[\rho]=\frac{\tr[\Gamma_C]}{3}=\frac{F_Q[\rho;\hat J_x]+F_Q[\rho;\hat J_y]+F_Q[\rho;\hat J_z]}{3}.
\ee
The sum of three Fisher informations 
for the phase generators $\hat J_x$, $\hat J_y$, and $\hat J_z$ on the right 
hand side appeared already in Refs~\cite{WieczorekPrep,WieczorekPhD09}
as a criterion for entanglement.
We would like to determine bounds on $\overline F_Q$ for $k$-producible
states in analogy to the bounds that we found for $F_Q$. 
We directly state the results and derive them afterwards.

\noindent
{\bf Observation 2 (${\bf \overline F_Q^{k+1}}$ criterion)}.
{\em For $k$-producible states,
the average quantum Fisher information defined in Eq.~(\ref{eq:Fav2}) 
is bounded by
\be \label{Fav_k_ent}
	\overline F_Q[\rho_{k-{\rm prod}}] \le \frac{1}{3}[s(k^2+2k-\delta_{k,1})+r^2+2r-\delta_{r,1}],
\ee
where $s=\lfloor \frac{N}{k}\rfloor$, $r=N-s k$ and $\delta$ is the Kronecker delta. 
Hence a violation of the bound~(\ref{Fav_k_ent}) proves
$(k+1)$-particle entanglement.
For separable states, corresponding to $k=1$,
the bound becomes 
\be
	\label{Fav_sep}
	\overline F_Q[\rho_{\rm sep}] \le \frac{2}{3}N.
\ee
The maximal value for any quantum state is 
given by 
\be\label{Fav_bound}
	\overline F_Q\le \frac{1}{3}[N^2 + 2N].
\ee
}

{\em Proof.}
Let us first prove Eq.~(\ref{Fav_bound}).
Since $\overline F_Q$ can be written as the sum of three quantum Fisher informations,
it is also convex in the states. Therefore, the maximum is again reached
for pure states. Hence $\overline F_Q\le 
\frac{4}{3}\max_{\ket{\psi}}[\mean{\vec{\hat J}^2}_{\ket{\psi}}-
\vert \mean{\vec{\hat J}}_{\ket{\psi}} \vert^2]\le \frac{4}{3} j(j+1)$, 
where $\mean{\vec{\hat J}^2} \equiv \mean{\vec{\hat J} \cdot \vec{\hat J}} = 
\mean{\hat J_x^2}+\mean{\hat J_y^2}+\mean{\hat J_z^2}$
and $\vert\mean{\vec{\hat J}}\vert^2 = \mean{\hat J_x}^2+\mean{\hat J_y}^2+\mean{\hat J_z}^2$.
This leads to Eq.~(\ref{Fav_bound}) because $\vert \mean{\vec{\hat J}}_{\ket{\psi}} \vert^2\ge 0$ 
and 
\be\label{Jtot}
	\mean{\vec{\hat J}^2}_{\ket{\psi}}\le j(j+1)
\ee 
holds in general,
while equality is reached by the symmetric states of $N$ particles
\cite{nota_Dicke}.

For the $k$-producible pure state 
$\ket{\psi_{k-{\rm prod}}}=\bigotimes_{l=1}^M\ket{\psi_l}$, 
the average quantum Fisher information is given by
$\overline F_Q[\ket{\psi_{k-{\rm prod}}}] = 
\frac{4}{3}\sum_{l=1}^M [\mean{\vec{\hat J}_l^2}_{\ket{\psi_l}}-
\vert \mean{\vec{\hat J}_l}_{\ket{\psi_l}}\vert^2]\le \frac{1}{3}\sum_{l=1}^M[N_l^2+2N_l
-4\vert\mean{\vec{\hat J}_l}_{\ket{\psi_l}}\vert^2]$, where $\vec{\hat J}_l$ is the 
vector of collective spin operators acting on the particles 
contained in state $\ket{\psi_l}$. The inequality is due to Eq.~(\ref{Jtot}).
In the same way as it was for $F_Q$, in order to maximize the bound it is advantageous 
to increase the $N_l$ as much as possible. This is true even though if $N_l=1$
then $\overline F_Q$ is reduced by $\frac{1}{3}$ since 
$\vert\mean{\vec{\hat J}}_{\ket{\psi_l}}\vert^2=\frac{1}{4}$ in this case.
For $k\in[1,N]$, we obtain the bound~(\ref{Fav_k_ent}),
where 
$s=\lfloor \frac{N}{k}\rfloor$ and $r=N-s k$ as above,
and we obtain Eq.~(\ref{Fav_sep}) for $k=1$.
\proofend

\begin{figure}[!h]
\begin{center}
\includegraphics[scale=0.58]{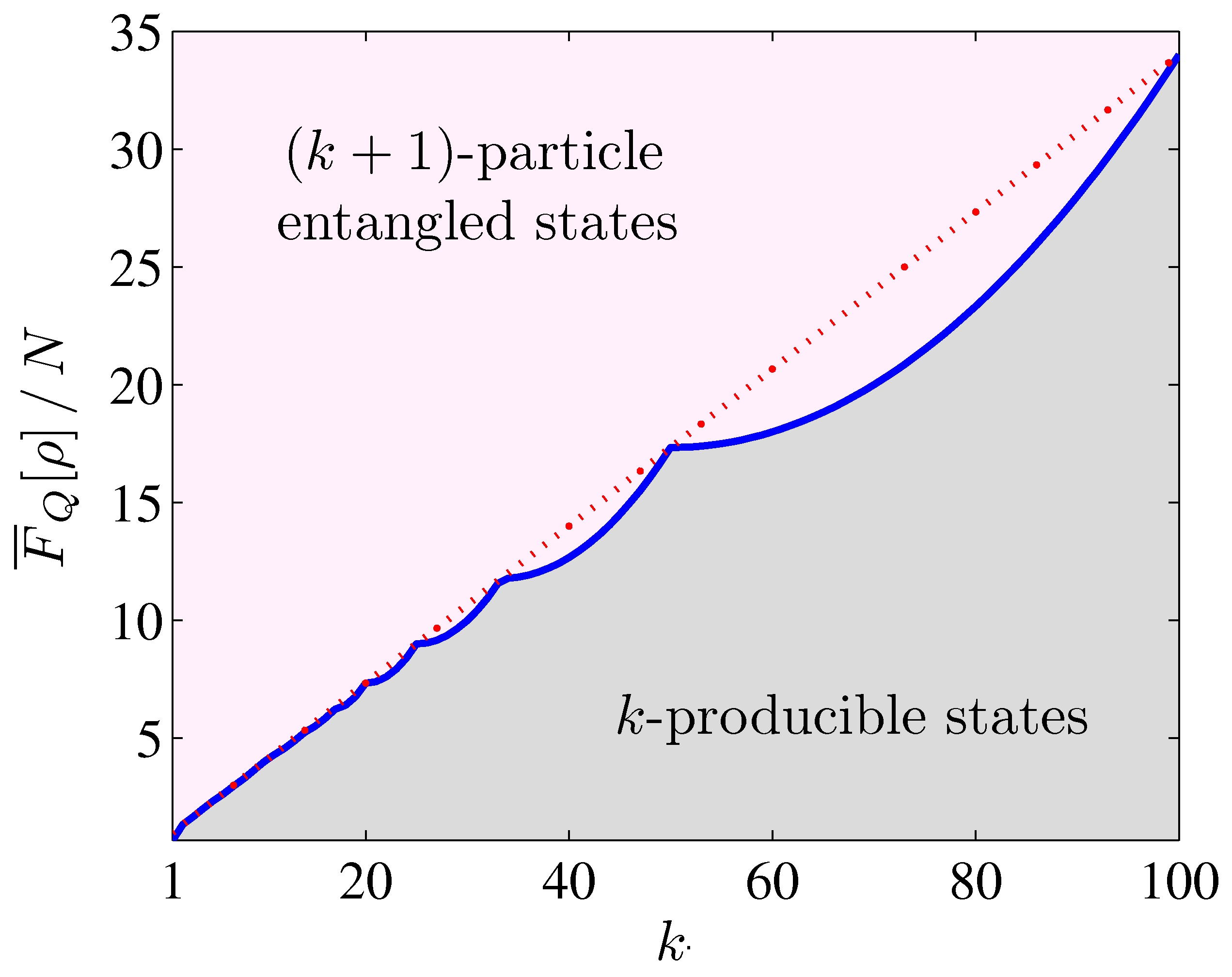}
\end{center}
\caption{\small{$\mathbf{\overline F_Q^{k+1}}$ {\bf criterion:} 
The solid line shows the bound in Eq.~(\ref{Fav_k_ent}) [$\overline F_Q^{k+1}$-criterion] 
as a function of $k$. For comparison, the function $N(k+2)/3$ is plotted (dotted line). 
Here $N=100$.}}
\label{Fig2} 
\end{figure}

The bound in Eq.~(\ref{Fav_k_ent}) is shown in Fig.~\ref{Fig2} as a function of $k$.
Let us note that the bound for $k=N-1$ is 
\be
	\label{Fav_N-1_ent}
	\overline F_Q[\rho_{(N-1)-{\rm prod}}] \le \frac{1}{3}[N^2+1].
\ee
Again, the bounds for a given $k$ 
are saturated by using $s$ GHZ states of $k$ 
particles and one GHZ state of $r$ particles.
However, as we shall discuss presently, these
states are not uniquely saturating the bounds,
in contrast to what happens in the case of the
$F_Q^{k+1}$ criterion.

\subsection{${\bf F_Q^{k+1}}$ criterion vs ${\bf \overline F_Q^{k+1}}$ criterion}
\label{ssec:FQ_vs_Fav}

\begin{table}
\begin{tabular}{c|c|c|c}
State & $\Gamma_C$ & $F_Q^{\rm max}$ & $\overline F_Q$\\
\hline
$\ket{1}^{\otimes N}$ & $\diag(N,N,0)$ & $N$ & $\frac{2}{3}N$ \\[1mm]
$\ket{{\rm GHZ}_N}$ & $\diag(N,N,N^2)$ & $N^2$ & $\frac{1}{3}(N^2+2N)$ \\[1mm]
$\ket{D_N^{(N/2)}}$ & $\frac{1}{2}(N^2+2N)\diag(1,1,0)$ & $\frac{1}{2}(N^2+2N)$ & $\frac{1}{3}(N^2+2N)$\\[1mm]
\hline \hline
\end{tabular}
\caption{\label{tab:FQ_vs_FQav}
Comparison of the maximal values of $F_Q$ and $\overline F_Q$ for three different input states.
}
\end{table}

To start the comparison, let us first discuss states with extremal 
values for the criteria. In particular, we consider the cases $k=1$,
where the criteria detect any kind of entanglement, and $k=N-1$,
where the criteria detect genuine multiparticle entanglement.
The states we use to illustrate the criteria are the $N$-particle 
GHZ state from Eq.~(\ref{eq:GHZ}), the
fully separable state $\ket{1}^{\otimes N}$, and the 
Dicke state with $N/2$ excitations \cite{nota_Dicke} 
\be\label{eq:Dicke}
\ket{D_N^{(N/2)}}={\cal S}(\ket{0}^{\otimes N/2}\otimes\ket{1}^{\otimes N/2}),
\ee
known as twin-Fock state for indistinguishable particles \cite{HollandPRL93}.
In Tab.~\ref{tab:FQ_vs_FQav}, we list the $\Gamma_C$ matrices 
for these states for all $N$. Since all states are symmetric under the
exchange of any two particles, we can directly read off the optimal
values of $F_Q$ and $\overline F_Q$ from these matrices \cite{HyllusPRA10}.
For the pure separable state, $F_Q[\ket{1}^{\otimes N};\hat J_{\vec n}]=N$ 
for any direction ${\vec n}$ in the $x-y$ plane because $\ket{1}$ 
is an eigenstate of $\hat\sigma_z$. Hence this state saturates
the bound for separable states both for $F_Q$ and for $\overline F_Q$.
As noted before, the GHZ state maximizes both $F_Q$ and $\overline F_Q$.
The Dicke state $\ket{D_N^{(N/2)}}$ has a $N^2$-scaling in 
$F_Q$ as the GHZ state with a prefactor $\frac{1}{2}$,
therefore, it does not saturate the maximum value $F_Q = N^2$. However, it 
saturates the maximal value of $\overline F_Q$ from Eq.~(\ref{Fav_bound}). 
In fact, the criterion $F_Q^N$ detects $\ket{D_N^{(N/2)}}$ as $N$-particle entangled
if $N\le 5$ only, while the criterion $\overline F_Q^N$ detects the state as $N$-particle entangled for any 
value of $N$.
Hence $\overline F_Q$ is not uniquely saturated by the GHZ state 
as $F_Q$. We will use this fact in the proof of the following 
Observation which shows that the two criteria in general detect
strictly different sets of states.

\noindent
{\bf Observation 3.} 
{\em (a) For all pairs $(k,N)$ with $k<N$, the $F_Q^{k+1}$
criterion detects the entanglement of some states 
for which the $\overline F_Q^{k+1}$ does not detect entanglement.
(b) For all pairs $(k,N)$ with $2<k<N$, the $\overline F_Q^{k+1}$ 
criterion detects the entanglement of some states 
for which the $F_Q^{k+1}$ criterion does not detect entanglement.
}

The proof can be found in the Appendix. Part (b) of Observation 3 can
be extended also to cases where $k=1<N$ and $k=2<N$, as shown in 
Sections \ref{sec:ExtObs3} and \ref{sec:BE} below.

\section{Examples}
\label{sec:examples}

We will now turn to illustrate the strength of the described criteria for their utilization in entanglement detection and in quantum metrology applications. To this end, we evaluate the criteria for different sets of states. We will first consider an actual experimental setting of different types of entangled four-qubit states. Secondly, we will consider various three-qubit entangled states including bound entangled states. We will compare different means to detect their entanglement by computing the amount of detected states. Finally, we construct 
an example extending Observation 3 before 
we examine two families of bound entangled states.

\subsection{Experimental GHZ and Dicke states}

We start by applying the above criteria to entangled states of $N=4$ 
photonic qubits produced experimentally by parametric downconversion 
from the Refs~\cite{WieczorekPRL08,KrischekPRL11}. The qubits
are encoded in the polarization with $\ket{0}\equiv\ket{H}$
and $\ket{1}\equiv\ket{V}$, where $H$ stands for horizontal
and $V$ for vertical polarization. In Ref.~\cite{WieczorekPRL08},
a large family of entangled states of $N=4$ qubits has 
been produced. We will investigate the data of the state 
$\frac{1}{\sqrt{2}}(\ket{0011}+\ket{1100})$, which can be converted
to a GHZ state [cf. Eq.~(\ref{eq:GHZ})] by flipping the state of the last two qubits,
and the state $\ket{\psi^+}\otimes\ket{\psi^+}$, where 
$\ket{\psi^+}=\frac{1}{\sqrt{2}}(\ket{01}+\ket{10})=\ket{D_2^{(1)}}$.
Hence this state is a product of two-particle Dicke states \cite{nota_Dicke}.
Note that by flipping the state of the second and of the fourth qubit,
this state can be transformed into $\ket{{\rm GHZ}_2}\otimes \ket{{\rm GHZ}_2}$.
Finally, we will also use the data of Ref.~\cite{KrischekPRL11}, where
the Dicke state $\ket{D_4^{(2)}}$ has been produced.
The states where observed with fidelities $\mathcal{F}_{{\rm GHZ_4}}=0.8303 \pm 0.0080$,
$\mathcal{F}_{(D_2^{(1)})^{\otimes 2}}=0.9255 \pm 0.0091$ \cite{WieczorekPRL08}, and
$\mathcal{F}_{D_4^{(2)}}= 0.8872 \pm  0.0055$ \cite{KrischekPRL11}. For comparison,
the data of the separable state $\ket{+}^{\otimes 4}$
measured in Ref.~\cite{KrischekPRL11} 
is used, which was observed with a fidelity $\mathcal{F}_{\ket{+}^{\otimes 4}}=0.9859\pm 0.0062$.
Here, $\ket{+}=\frac{1}{\sqrt{2}}(\ket{0}+\ket{1})$.
 
\begin{figure*}[!t]
\begin{center}
\includegraphics[scale=0.65]{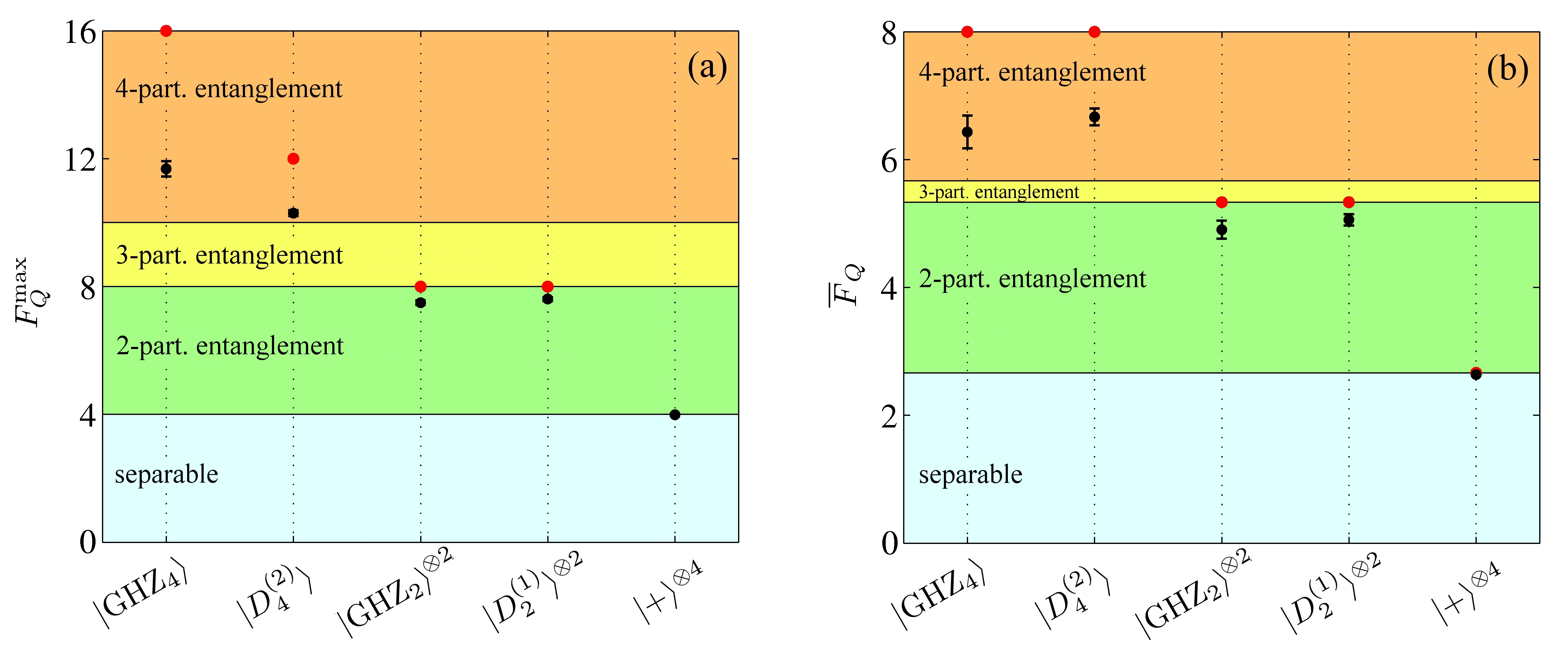}
\end{center}
\caption{\small{Black dots are the values of $F_Q^{\rm max}$ [in panel (a)] and of $\overline F_Q$ 
[in panel (b)] for states generated 
experimentally, calculated from the experimental density matrices. 
Error bars are calculated with a parametric bootstrap
method \cite{EfronBook} (see also the supplementary material of Ref.~\cite{KrischekPRL11}). 
The red dots are ideal values. 
More explicitly, we have:
$F_Q^{\rm max} = 11.681\pm 0.238$ and $3\overline F_Q =19.296\pm 0.256$ for the $\ket{{\rm GHZ}_4}$ state;
$F_Q^{\rm max} = 10.291\pm 0.094$ and $3\overline F_Q = 20.004\pm 0.131$ for the $\ket{D_4^{(2)}}$ state;
$F_Q^{\rm max} = 7.495\pm 0.070$ and $3\overline F_Q = 14.713\pm 0.141$ for the $\ket{{\rm GHZ}_2}^{\otimes 2}$ state;
$F_Q^{\rm max} =7.612\pm 0.058$ and $3\overline F_Q =15.174\pm 0.089$ for the $\ket{D_2^{(1)}}^{\otimes 2}$ state;
$F_Q^{\rm max} =4.002\pm 0.025$ and $3\overline F_Q =7.902\pm 0.015$ for the separable 
state $\ket{+}^{\otimes 4}$.
In panel (a) and (b), the vertical lines are bounds for the $F_Q^{k+1}$ 
and $\overline F_Q^{k+1}$ criteria, respectively. }}
\label{Fig3} 
\end{figure*}

The optimized quantum Fisher information $F_Q^{\rm max}$ and 
$\overline F_Q$ for the different states 
are calculated from the measured density matrix.
We compare the experimental results with the ideal cases 
and with the bounds on $k$-producible states 
from Observations 1 and 2 for $N=4$.
In order to do so, we apply the bit-flips mentioned above to the experimental 
data where necessary. 
The results are shown in Fig.~(\ref{Fig3}).
For the $N=4$ GHZ and Dicke states, $4$-particle entanglement is proven
with a high statistical significance by $\overline F_Q$. 
In particular, for the Dicke
state, the statistical significance for the proof of $4$-particle
entanglement from $F_Q^{\rm max}$ is much lower. This is a consequence
of the fact  that the ideal Dicke states reaches 
the maximal value of $\overline F_Q$ for any $N$, while the deviation of
$F_Q^{\rm max}$ from the maximal value increases with increasing $N$
[cf. Tab.~\ref{tab:FQ_vs_FQav}]. 
The very high fidelity of the experimental product of two $N=2$ Dicke states
is reflected in the fact that $F_Q^{\rm max}$ and $\overline F_Q$ nearly 
reach the optimal values for the states $\ket{D_2^{(1)}}^{\otimes 2}$
and $\ket{{\rm GHZ}_2}^{\otimes 2}$, 
and entanglement is 
clearly proven, while the bounds for $2$-particle entangled
states are not violated. 

As a final remark, we would like to point out that 
the multiparticle entanglement of the states could be proved
with less experimental effort and a generally larger statistical
significance by witness operators \cite{WieczorekPRL08}. 
However, in this case this does not give any direct information
about the usefulness for a given task, in particular for phase estimation
[See Ref.~\cite{KrischekPRL11}
for a detailed comparison of the $F_Q$ criteria with a witness
operator for the state $\ket{D_4^{(2)}}$].

\subsection{Pure states of 3 particles}

In order to get an impression of the strength of the criteria, 
we randomly choose a three-qubit state $|\psi\rangle$ and analyze it using various criteria.
First, we evaluate the criteria $F_Q^2$ and $\overline F_Q^2$ which detect entanglement. 
Further, we compare several criteria detecting multiparticle entanglement:
(i) the entanglement witness ${\cal W}=\frac{1}{2}\openone-\proj{{\rm GHZ}}$,
which has a positive expectation value for all 2-particle entangled states \cite{WITNESS},
(ii) the density matrix element condition (DME) which states that 
\be\label{DME}
	|\rho_{18}| \leq \sqrt{\rho_{22}\rho_{77}}+\sqrt{\rho_{33}\rho_{66}}+\sqrt{\rho_{44}\rho_{55}}
\ee
for all 2-entangled states ($\rho_{ij}$ denote coefficients of a given density matrix $\rho = |\psi\rangle\langle\psi|$) \cite{GuehneNJP10}, and (iii) the multiparticle criteria $F_Q^3$ and $\overline F_Q^3$.

To generate a random pure state \cite{ZYCZKOWSKI}, we take a vector of a random unitary matrix
distributed according to the Haar measure on U(8): 
\begin{eqnarray}
|\psi\rangle &=& (\cos{\alpha_7}, \cos{\alpha_6}\sin{\alpha_7}e^{i \phi_7}, \cos{\alpha_5}\sin{\alpha_6}\sin{\alpha_7}e^{i \phi_6}, 
\nonumber \\
&\ldots&, \sin{\alpha_1}\cdots \sin{\alpha_7}e^{i \phi_1}),
\end{eqnarray}
where $\alpha_i \in [0,\pi/2]$ and $\phi_k=[0,2\pi)$. The parameters are drawn with the probability densities: $P(\alpha_i)=i \sin(2\alpha_i)(\sin \alpha_i)^{2i-2}$ and $P(\phi_i)=1/2\pi$. The calculations were performed for a set of $10^6$ states. The results are presented in Tab.~{\ref{t-rand}}.
The averaged criteria seem to detect more states in general. It is surprising that
the witness condition detects nearly as many states as the criteria $F_Q^3$ and $\overline F_Q^3$.
This may be an artifact of the small $N$ we chose.

\begin{table}
\begin{tabular}{l | c } 
\hline
Criterion & detected $2$-ent. [\%] \\
\hline
$F_Q^2$  & 94.32  \\
$\overline F_Q^2$  & 98.38  \\
\hline
\hline
Criterion & detected $3$-ent. [\%] \\
\hline
${\cal W}$ & 18.99 \\
DME  & 80.63 \\
DME'  & 82.61\\
$F_Q^3$  & 22.93 \\
$\overline F_Q^3$  & 27.99\\
\hline\hline
\end{tabular}
\caption{\label{t-rand} 
Percentage of detected $2$-particle and $3$-particle entangled pure three-qubit states.
See text for details. 
DME' denotes the whole family of DME conditions, 
which is obtained by permuting the qubits of the 
state.}
\end{table}

\subsection{GHZ-diagonal states}

The DME criterion~(\ref{DME}) and the criteria obtained thereof by permutations of the qubits
completely characterize the GHZ-diagonal states of three qubits \cite{GuehneNJP10}, 
which can be written as
\be\label{GHZdiag}
	\frac{1}{{\cal N}}
	\left(
	\begin{array}{cccccccc}
		\lambda_1 & 0 & 0 & 0 & 0 & 0 & 0 & \mu_1 \\
		0 & \lambda_2 & 0 & 0 & 0 &  0 & \mu_2 & 0 \\
		0 & 0 & \lambda_3 & 0 & 0 &  \mu_3 & 0 & 0 \\
		0 & 0 & 0 & \lambda_4 & \mu_4 & 0 & 0 & 0 \\
		0 & 0 & 0 & \mu_4 & \lambda_5 & 0 & 0 & 0 \\
		0 & 0 & \mu_3 & 0 & 0 &  \lambda_6 & 0 & 0 \\
		0 & \mu_2 & 0 & 0 & 0 &  0 & \lambda_7 & 0 \\
		\mu_1 & 0 & 0 & 0 & 0 & 0 & 0 & \lambda_8
	\end{array}
	\right)
\ee
with real coefficients $\lambda_i$ and $\mu_i$,
where ${\cal N}$ is a normalization factor.
If $\lambda_i=\lambda_{9-i}$ for $i=5,6,7,8$, then
these states are diagonal in the GHZ-basis 
$\ket{\psi_{l_1 l_2}^\pm}=\frac{1}{\sqrt{2}}(\ket{0l_1 l_2}\pm\ket{1\bar l_1\bar l_2})$,
where $l_1$ and $l_2$ are equal to $0$ or $1$, and $\bar 1=0$ and 
$\bar 0 = 1$. 
We generated $10^6$ random states of this form violating Eq.~(\ref{DME}) directly, 
which states $|\mu_1|\le \lambda_2+\lambda_3+\lambda_4$ in this case.
The results are shown in Tab.~\ref{tab2} in the middle column. 
Then, we generated again $10^6$ states violating Eq.~(\ref{DME}) or its other forms
obtained by permuting the qubits. The results are shown in
the right column of Tab.~\ref{tab2}. The witness criterion detects
significantly more states than the criteria based on the Fisher information.
Contrary to the case of pure states, the $F_Q^3$ criterion detects
more states than $\overline F_Q^3$ in this case.
Note that the percentage of detected states reduces significantly 
for all criteria in the DME' case. The reason is that all criteria 
work best for the symmetric GHZ state $\frac{1}{\sqrt{2}}(\ket{000}+\ket{111})$, 
which has the highest weight in the state if only condition~(\ref{DME}) is 
used \cite{GuehneNJP10}.

\begin{table}
\begin{tabular}{l | c | c } 
Criterion & detected DME [\%] & detected DME' [\%]\\
\hline
${\cal W}$ &  50.56 & 12.27 \\
$F_Q^3$  & 19.45 & 4.77 \\
$\overline F_Q^3$ & 13.14 & 3.25 \\
\hline\hline
\end{tabular}
\caption{\label{tab2} Percentage of GHZ-diagonal 3-particle entangled states 
which are detected by the entanglement witness, the criterion $F_Q^3$ and the crition $\overline F_Q^3$.
In the middle column, only states violating the DME condition~(\ref{DME})
have been generated, while in the last column, also states violating
any of the other DME conditions obtained by permutations of the particles
have been generated.
}
\end{table}

The family of states~(\ref{GHZdiag}) also comprises bound entangled states 
if $\lambda_1=\lambda_8=\mu_1=1$ and $\lambda_7=1/\lambda_2$,
$\lambda_6=1/\lambda_3$, $\lambda_5=1/\lambda_4$, and $\mu_2=\mu_3=\mu_4=0$,
as long as $\lambda_2\lambda_3\neq\lambda_4$. Then the states have 
a positive partial transpose (PPT) \cite{PPT} for any bipartition of the
three particles while still being entangled \cite{AcinPRL01}. 
It follows that the state
cannot be distilled to a GHZ state \cite{HorodeckiPRL98,DuerPRL99}.
We generated again $10^6$ random states of this class and applied
$F_Q^2$ and $\overline F_Q^2$, but neither criterion detected any of these
states. However, we will see presently that $\overline F_Q$ is in fact 
able to detect bound entanglement.

\subsection{Extension of the Observation 3 for $N=4$}
\label{sec:ExtObs3}

Observation 3 (b) can be extended to pairs $(k,N)$
where $1\le k<N$. We now construct an explicit example
for the cases $N=4$ and $k=1,2$. 
The basic idea is to use states 
with the property $\Gamma_C=c_N\openone$ 
which are extremal in the sense that
they saturate the inequality 
$\max_{\vec n}F_Q[\rho;\hat J_{\vec n}] \ge \overline F_Q[\rho]$.
Hence they provide the minimal $F_Q^{\rm max}$ compared to $\overline F_Q$. 
One way of constructing such states is by 
considering a symmetric state $\ket{\psi_S}=\sum_\mu \gamma_\mu\ket{j,\mu}$ \cite{nota_Dicke}, 
and by choosing the $\gamma_\mu$ such that $\mean{\vec{\hat J}}=0$ and 
$\mean{\hat J_x^2}=\mean{\hat J_y^2}=\mean{\hat J_z^2}$.
If $\gamma_\mu\neq 0$ and $\gamma_{\mu'}\neq 0$ only if
$|\mu-\mu'|>2$ then
$\mean{\hat J_x}=\mean{\hat J_y}=\mean{\hat J_i\hat J_j}=0$ for $(i,j)=(x,y)$,
$(i,j)=(x,z)$, and $(i,j)=(y,z)$. For $N=4$, all the conditions above 
are fulfilled by the 
states
\be 
\ket{\psi_S^4\pm}=\sqrt{\frac{1}{3}}\ket{2,\pm 2}+\sqrt{\frac{2}{3}}\ket{2,\mp 1}, 
\ee
leading
to $\Gamma_C=8\openone$, and hence $F_Q^{\rm max}=\overline F_Q=8$.
With this state,  $\overline F_Q$ reaches the maximal value possible for $N=4$, cf. 
Eq.~(\ref{Fav_bound}), while $F_Q^{\rm max}$ saturates the bound 
of Eq.~(\ref{FQ_class}) for $k=2$. This provides the example for
Observation 3 (b) for $(N,k)=(4,2)$. 
If we mix $\ket{\psi_S^4\pm}$ with the identity,
then using Eq.~(\ref{GammaC_rho_p}) from the Appendix it can be shown that for 
$p^*=\frac{7}{32}(1+\sqrt{113}/7)$, we obtain $\Gamma_C[\rho(p^*)]=4\openone$. 
Hence $\rho(p^*)$ saturates the $F_Q^{k+1}$ criterion but
violates the $\overline F_Q^{k+1}$ criterion for $k=1$.
This provides the example for
Observation 3 (b) for $(N,k)=(4,1)$. 

Note that the state $\ket{\psi_S^4-}$ has appeared also in other
contexts \cite{GuehnePR09}. For instance, it is the most non-classical state for total 
spin $j=2$ \cite{GiraudNJP10}, and it is a maximally entangled state of 4 qubits for 
multipartite entanglement measures  based on anti-linear operators and combs 
\cite{OsterlohPRA05}. Finally, symmetric 
states with $\Gamma_C\propto\openone$ have the highest sensitivity to small
misalignments of Cartesian reference frames \cite{KolenderskiPRA08}. 
The quantity to be optimized in the derivations is $3\overline F_Q$.
For $N=4$, again the state $\ket{\psi_S^4-}$ is optimal, and several other
examples of symmetric states with $\Gamma_C\propto\openone$ 
for even $N$ have been found in Ref.~\cite{KolenderskiPRA08}. 

The bound entangled D{\"u}r and Smolin states
considered in Sec.~\ref{sec:BE} 
below provide further examples, for $k=1$ and any $N$.

\subsection{Detecting bound entangled states}
\label{sec:BE}

We consider two families of states where the state has 
a PPT with respect to some bipartitions, but not with respect to others. 
Due to the PPT bipartitions it is not possible to distill
these states to a GHZ state non{\blue e}theless \cite{DuerPRL99}.
Both families of states provide examples for situations
where the $\overline F_Q^{k+1}$ criterion detects states which
the $F_Q^{k+1}$ does not detect for $k=1$ and for any value
of $N$. This extends the results summarized in Observation 3 from
Sec.~\ref{ssec:FQ_vs_Fav}.

\subsubsection{D\"ur states}

Interestingly, the $\overline F_Q^{2}$ criterion 
(\ref{Fav_sep}) can reveal entanglement of a bound entangled state introduced
by D\"ur \cite{DUR}:
\begin{equation}\label{eq:Duer}
\rho_{\rm D\ddot{u}r}^{(N)} = \frac{1}{N+1} \Big( \ketbra{{\rm GHZ}_\varphi} + \frac{1}{2} \sum_{l=1}^N (P_l + \bar P_l) \Big),
\end{equation}
with $|{\rm GHZ}_\varphi \rangle = \frac{1}{\sqrt{2}} \left[ \ket{0}^{\otimes N} + e^{i \varphi} \ket{1}^{\otimes N} \right]$, where $\varphi$ is an arbitrary phase. 
We will consider $\varphi=0$ in the following.
Further, $P_l$ is the projector on the state 
$\ket{0}^{\otimes l-1}\otimes\ket{1}\otimes\ket{0}^{\otimes N-l}\equiv \ket{1_l}$
and $\bar P_l$ is obtained from $P_l$ by exchanging $0\leftrightarrow 1$.

We can directly state the eigenstates and eigenvalues.
The state $\ket{{\rm GHZ}_0}$ is an eigenstate 
with eigenvalue $\frac{1}{N+1}$ and the states $\ket{1_l}$ and $\ket{0_l}$ 
are eigenstates with eigenvalue $\frac{1}{2(N+1)}$. The kernel is spanned 
by the state $\ket{{\rm GHZ}_\pi}$ and by all states of the form 
$\ket{n_{\cal P}}\equiv{\cal P}(\ket{0}^{\otimes n}\otimes\ket{1}^{\otimes N-n})$, where ${\cal P}$
is a permutation of the qubits and $n=2,3,...,N-2$. 
Now we can compute the elements of the correlation matrix $\Gamma_C$ 
using Eq.~(\ref{eq:GammaCentries}). 
The nonvanishing factors $\bra{l}\hat J_i\ket{l'}$
are given in Tab.~\ref{tab:DuerState}.
We obtain 
\be 
	\Gamma_C=N \diag\Big(\frac{3N-1}{3N+3},\frac{3N-1}{3N+3},\frac{N}{N+1}\Big).
\ee
The matrix $\Gamma_C$ is diagonal
because the factors $\bra{l}\hat J_{x,z}\ket{l'}$ are real while the 
factors $\bra{l}\hat J_{y}\ket{l'}$ are imaginary, and since $\bra{l}\hat J_{x}\ket{l'}$
vanishes for the eigenstates where $\bra{l}\hat J_{z}\ket{l'}\neq 0$ and vice versa.
We observe that $F_Q^{\rm max} < N$ for all $N$ while 
\be
	\overline F_Q=\frac{9N-2}{9N+9} N>\frac{2}{3}N
\ee 
for all $N$. 

\begin{table}
\begin{tabular}{l | c | c} 
Factor & value &  multiplicity\\
\hline
$\bra{{\rm GHZ_0}}\hat J_x\ket{1_l}$ & $\frac{1}{\sqrt{8}}$ &  N \\
$\bra{{\rm GHZ_0}}\hat J_x\ket{0_l}$ & $\frac{1}{\sqrt{8}}$ &  N \\
$\bra{{\rm GHZ_\pi}}\hat J_x\ket{1_l}$ & $\frac{1}{\sqrt{8}}$  & N\\
$\bra{{\rm GHZ_\pi}}\hat J_x\ket{0_l}$ & $-\frac{1}{\sqrt{8}}$  & N\\
$\bra{1_l}\hat J_x\ket{(N-2)_{\cal P}}$ & $\frac{1}{2}$ &  N(N-1) \\
$\bra{0_l}\hat J_x\ket{2_{\cal P}}$ & $\frac{1}{2}$ &   N(N-1) \\
\hline
$\bra{{\rm GHZ_0}}\hat J_y\ket{1_l}$ & $-\frac{i}{\sqrt{8}}$  & N\\
$\bra{{\rm GHZ_0}}\hat J_y\ket{0_l}$ & $\frac{i}{\sqrt{8}}$  & N\\
$\bra{{\rm GHZ_\pi}}\hat J_y\ket{1_l}$ & $-\frac{i}{\sqrt{8}}$  & N\\
$\bra{{\rm GHZ_\pi}}\hat J_y\ket{0_l}$ & $-\frac{i}{\sqrt{8}}$  & N\\
$\bra{1_l}\hat J_y\ket{(N-2)_{\cal P}}$ & $-\frac{i}{8}$ &  N(N-1)\\
$\bra{0_l}\hat J_y\ket{2_{\cal P}}$ & $-\frac{i}{2}$ & N(N-1)\\
\hline
$\bra{{\rm GHZ}_\pi}\hat J_z\ket{{\rm GHZ}_0}$ & $\frac{N}{2}$ &  1\\
\hline\hline
\end{tabular}
\caption{\label{tab:DuerState} 
Nonvanishing factors contributing to $\Gamma_C$ for the D{\"u}r states [Eq.~(\ref{eq:Duer})].
The multiplicity is the number of occurences.
}
\end{table}

Hence, the $F_Q^{2}$ criterion does not detect the 
entanglement in any of these cases, cf. Eq.~(\ref{F_Q_bound_sep}).
Therefore, these states represent an example of 
Observation 3 b) for $k=1$ and any $N$.
In conclusion, the states are not useful for sub shot-noise 
interferometry for any direction ${\vec n}$,
even though they are more useful than separable states when averaging 
over all directions. 

\subsubsection{Generalized Smolin states}

As a second example, we consider the generalized $N=2n$-qubit Smolin state \cite{SMOLIN}
\be \label{eq:Smolin}
\rho^{(N)}_{\rm Smolin} = \frac{1}{2^{N}} (\openone + (-1)^n \sum_{i=1}^3 \sigma_i^{\otimes N}),
\ee
which can be written a mixture of $2n$-qubit GHZ-type states,
\be
\rho^{(N)}_{\rm Smolin} = \frac{1}{2^{N-2}} \sum_{\sum_j i_j\ {\rm even/odd}} 
\ketbra{{\rm GHZ}_0^{i_1...i_{N}}}, 
\ee 
where 
$\ket{{\rm GHZ}_{\varphi}^{i_1...i_{N}}} = 
\frac{1}{\sqrt{2}}[\ket{i_1,i_2,...,i_{N}}
+e^{i\varphi} \ket{\bar i_1,\bar i_2,...,\bar i_N}$. 
The index $i_j$ can take the values $0$ and $1$, and 
if $i_j=0$ then $\bar i_j=1$ and vice versa. 
For even $n$, then sum 
$\sum_{j=1}^N i_j\equiv N_1$ can take even values $\{0,2,...,n\}$, while if $n$ 
is odd, then the sum can take odd values
$\{1,3,...,n\}$.
The kernel of $\rho^{N}_{\rm Smolin}$ is spanned
by the states 
$\ket{{\rm GHZ}_\pi^{i_1...i_{N}}}$ for any set $\{i_j\}$
such that $N_1=0,1,...,n$,
and the states $\ket{{\rm GHZ}_0^{i_1...i_{N}}}$ with 
$N_1=1,3,...,n-1$ if $n$ is even and 
$N_1=0,2,...,n-1$ if $n$ is odd.

Now we can compute the elements of the correlation matrix $\Gamma_C$ 
using Eq.~(\ref{eq:GammaCentries}). 
The nonvanishing factors $\bra{l}\hat J_i\ket{l'}$
are given in Tab.~\ref{tab:SmolinState}.
We obtain 
\be 
	\Gamma_C=N\cdot\openone
\ee
for any even $N$.
The matrix $\Gamma_C$ is diagonal for the same reasons as in the 
previous case. We observe that $F_Q^{\rm max} = N$ for all $N$ while 
\be
	\overline F_Q= N>\frac{2}{3}N
\ee 
for all $N$. Therefore, these states represent an example of 
Observation 3 b) for $k=1$ and any {\rm even} $N$.

\begin{table}
\begin{tabular}{l | c | c} 
Factor & value &  multiplicity\\
\hline
$\bra{{\rm GHZ}_0^{i_1...i_N}} \hat J_z \ket{{\rm GHZ}_\pi^{i_1...i_N}}$ & $\frac{(N-2N_1)}{2}$ &  
${N\choose N_1}$ \\
$\bra{{\rm GHZ}_0^{i_1...i_r...i_N}} \hat J_{x} \ket{{\rm GHZ}^{i_1...\bar i_r...i_N}_{0}}$ & $\frac{1}{2}$ &  $N2^{N-2}$ \\
$\bra{{\rm GHZ}_0^{i_1...i_r...i_N}} \hat J_{y} \ket{{\rm GHZ}^{i_1...\bar i_r...i_N}_{\pi}}$ & $(-1)^{\bar i_r}\frac{i}{2}$ &  $N2^{N-2}$ \\
\end{tabular}
\caption{\label{tab:SmolinState} Nonvanishing factors contributing to $\Gamma_C$ for the Smolin states [Eq.~(\ref{eq:Smolin})].
The multiplicity is the number of occurences. 
}
\end{table}
Hence, similarly as previously, the $F_Q^{2}$ criterion does not detect the 
entanglement in any of these cases, cf. Eq.~(\ref{F_Q_bound_sep}), so 
the states are also not useful for sub shot-noise 
interferometry for any direction ${\vec n}$,
even though they are more useful than separable states when averaging 
over all directions.

\section{Conclusions and Outlook}
\label{sec:conclusions}

We have introduced two criteria based on the quantum Fisher information (QFI)
for the detection of entangled states of different multiparticle entanglement
classes, and consequently of their usefullness for sub shot-noise
phase estimation.
Our first criterion is obtained from $F_Q[\rho, \hat H_{\rm lin}]$, for general
linear operators of $N$ qubits.
Our second criterion is related to quantum Fisher information for collective spin operators, 
averaged over all directions on the Bloch sphere. 
Both sets of criteria can be easily evaluated for a given state $\rho$ of an arbitrary 
number of particles, even if the state is mixed.
We considered several examples,
showing in particular that the average quantum Fisher information can 
be used to detect bound entangled states. It remains an interesting
open question whether or not there exist bound entangled states 
which are detected by the quantum Fisher information, since 
this would imply that such states could be used for sub shot-noise
interferometry. 

{\em Acknowledgements}. We thank G. T{\'o}th for discussions. 
We acknowledge support of the EU program Q-ESSENCE (Contract No.248095),
the DFG-Cluster of Excellence MAP, and of the EU project QAP. 
W.L. is supported by the MNiSW Grant no. N202 208538 and by the
Foundation for Polish Science (KOLUMB program).
The collaboration is a part of a DAAD/MNiSWprogram.
W.W. and C.S. acknowledge support by QCCC of the Elite Network of Bavaria.
P.H. acknowledges financial support of the ERC Starting Grant GEDENTQOPT.
L.P. acknowledges financial support by MIUR through FIRB Project No. RBFR08H058.

{\em Note added}: 
Independently from our work, an article on the relationship 
between multiparticle entanglement and the Fisher information 
has appeared \cite{TothPRA12}.

\section*{Appendix -- Proof of Observation 3}
We consider states of the form 
\be\label{rho_p}
	\rho(p)=p\proj{\psi}+(1-p)\frac{\openone}{2^N},
\ee
mixtures of a pure state and the totally mixed state. 
It can be shown directly from Eq.~(\ref{eq:GammaCentries}) that 
\be\label{GammaC_rho_p}
	\Gamma_C[\rho(p)]=\gamma_{p,N}\Gamma_C[\ket{\psi}],\ \ \gamma_{p,N}=\frac{p^2 2^{N-1} }{p (2^{N-1}-1)+1}
\ee
holds. 
The criteria~(\ref{FQ_class}) and (\ref{Fav_k_ent})
can be rewritten as $\gamma_{p,N} \le \alpha_{N,k}$ 
and $\gamma_{p,N} \le \bar\alpha_{N,k}$, respectively,
where 
\be\label{alpha1}
	\alpha_{N,k}=\frac{sk^2+r^2}{F_Q[\ket{\psi}]}
\ee
and
\be\label{alpha2}
	\bar\alpha_{N,k}=\frac{s(k^2+2k-\delta_{k,1})+r^2-2r-\delta_{r,1}}{4\tr(\Gamma_C[\ket{\psi}]}.
\ee 
In order to violate the criteria, 
\begin{equation}
	p>x\cdot\frac{1-2^{1-N}}{2}\left[1+\sqrt{1+\frac{1}{x}\frac{2^{3-N}}{(1-2^{1-N})^2}}\right]
	\nonumber
\end{equation}
has to hold, where $x=\alpha_{N,k}$ or $x=\bar \alpha_{N,k}$. 
The right hand side is strictly monotonic increasing 
with $x$. If, for instance, $\alpha_{N,k}<\bar\alpha_{N,k}$,
then the $F_Q^{N}$ criterion detects the states as multiparticle 
entangled already for a smaller value of $p$ than the $\overline F_Q^{N}$ 
criterion. Therefore, we can prove the claim by comparing the $\alpha$
coefficients for different states $\ket{\psi}$. 
However, the minimal $x$ has to be such that
at least one criterion detects the state for $p\le 1$.

For $\ket{\psi}$ we employ the GHZ states from Eq.~(\ref{eq:GHZ})
and the Dicke states from Eq.~(\ref{eq:Dicke}). 
The results summarized in Tab.~\ref{tab:FQ_vs_FQav}
ensure the following: (i)
there will always be a $p\in (0,1]$
such that $F_Q^{k+1}$ and $\overline F_Q^{k+1}$ detect $\rho(p)$
when $\ket{\psi}=\ket{{\rm GHZ}_N}$, and (ii)
there will always be a $p\in (0,1]$
such that $\overline F_Q^{k+1}$ detects $\rho(p)$
when $\ket{\psi}=\ket{D_N^{(N/2)}}$.

Let us start with the GHZ states. 
We check whether or not $\alpha_{N,k} < \bar\alpha_{N,k}$ is fulfilled. This condition is equivalent to
\be\label{Obs3_a}
	2sk(N-k)-Ns\ \delta_{k,1} + 2r(N-r)-N\delta_{r,1}>0.
\ee
Checking explicitly the cases (i) $1<r<k$, (ii) $1=r<k$, (iii) $r=0, k>1$, and
(iv) $k=1$ it can be shown that Eq.~(\ref{Obs3_a}) is always fulfilled.
Hence for the family of states $\rho(p)$ from Eq.~(\ref{rho_p})
with $\ket{\psi}=\ket{{\rm GHZ}_N}$, for every $N$ and $1\le k<N$ the $F_Q^{k+1}$ criterion
detects always states in addition to the states that the $\overline F_Q^{k+1}$ criterion
detects. This proves part (a) of Observation 3.

Let us now consider the Dicke states
and check whether or not $\alpha_{N,k} > \bar \alpha_{N,k}$
is always fulfilled in this case. The condition 
is equivalent to
\be
	(sk^2+n\delta_{k,1}-2nk)+(r^2+\delta_{r,1}-2r)>0.
\ee
Again checking all the cases, it can be seen that this is fulfilled
for $k>2$ and any $r<k$. Hence in these cases the criterion $\overline F_Q^{k+1}$
detects states in addition to those that $F_Q^{k+1}$ detects.
In fact, the criterion $F_Q^{k+1}$ may not even detect any of the states of this 
family.
\proofend


\begin{thebibliography}{99}
\bibitem{nota_FQpure}
This can be seen as follows \cite{BraunsteinPRL94}.
In Eq.~(\ref{QFI}), $\hat H$ can be replaced 
by $\Delta\hat H=\hat H-\mean{\hat H}$ without changing the value 
of $F_Q$. Then we can bound
$F_Q[\rho;\hat H]=2\sum_{l,l'}(\lambda_{l}+\lambda_{l'}) \big[(\lambda_{l}-\lambda_{l'})/(\lambda_{l}+\lambda_{l'})\big]^2 \bra{l}\Delta \hat H\ket{l'}\bra{l'}\Delta \hat H\ket{l}
\le 4\sum_{l,l'}\lambda_l \bra{l}\Delta \hat H\ket{l'}\bra{l'}\Delta \hat H\ket{l}.$ 
An upper bound to this expression is reached when the sum is extended to all 
values of $l$ and $l'$. This implies that $F_Q\le 4\mean{(\Delta\hat H)^2}$.
Since for a pure state, $\lambda_l$ is equal to 1 for one $l$ only,
it follows that $F_Q = 4\mean{(\Delta\hat H)^2}$ in this case.
\bibitem{nota_basis}
The eigenstates of $\hat\sigma_z$ are defined by 
$\hat\sigma_z\ket{\alpha}=(-1)^{\alpha}\ket{\alpha}$,
where $\alpha=0,1$.
\bibitem{nota_Dicke} Let us introduce the Dicke states with $N_1$ excitations 
\cite{DickePR54,KieselPRL07},
which can be written as 
$\ket{D_N^{(N_1)}}={\cal S}(\ket{0}^{\otimes N-N_1}\otimes\ket{1}^{\otimes N_1})$ \cite{nota_basis},
where ${\cal S}$ symmetrizes and normalizes the states.
Hence they are fully symmetric and eigenstates $\ket{j,\mu}$ of $\hat J_z$,
where $j=\frac{N}{2}$ and $\mu=\frac{N-2N_1}{2}\in\{-\frac{N}{2},-\frac{N}{2}+1,...,\frac{N}{2}\}$.
\end{thebibliography}
\end{document}